 \documentclass[12pt,preprint]{aastex}

\pdfoutput=1
\usepackage{epstopdf}
\usepackage{placeins}
\usepackage{float}
\usepackage{subfigure}
\usepackage{amsmath}
\usepackage{textcomp}
\usepackage{gensymb}

\begin{document}

\title{Tilt Angle and Footpoint Separation of Small and Large Bipolar Sunspot Regions Observed with HMI}

\author{B. H. McClintock}
\affil{University of Southern Queensland, Toowoomba, 4350, Australia}
\email{u1049686@umail.usq.edu.au}

\and 

\author{A. A. Norton}
\affil{HEPL, Stanford University, Palo Alto, CA 94305, USA}
\email{aanorton@stanford.edu}


\begin{abstract}
We investigate bipolar sunspot regions and how tilt angle and footpoint separation vary during emergence and decay. The Helioseismic and Magnetic Imager on board the Solar Dynamic Observatory collects data at a higher cadence than historical records and allows for a detailed analysis of regions over their lifetimes. We sample the umbral tilt angle, footpoint separation, and umbral area of 235 bipolar sunspot regions in Helioseismic and Magnetic Imager - Debrecen Data (HMIDD) with an hourly cadence. We use the time when the umbral area peaks as time zero to distinguish between the emergence and decay periods of each region and we limit our analysis of tilt and separation behavior over time to within $\pm 96$ hours of time zero.  Tilt angle evolution is distinctly different for regions with small ($\approx 30$ MSH), midsize ($\approx 50$ MSH), and large ($\approx 110$ MSH) maximum umbral areas, with 45 and 90 MSH being useful divisions in separating the groups.  At the peak umbral area, we determine median tilt angles for small (7.6\degr), midsize (5.9\degr) and large (9.3\degr) regions.  Within $\pm 48$ hours of the time of peak umbral area, large regions steadily increase in tilt angle, midsize regions are nearly constant, and small regions show evidence of negative tilt during emergence. A period of growth in footpoint separation occurs over a 72-hr period for al of thel regions from roughly 40 to 70 Mm. The smallest bipoles ($<9$ MSH) are outliers in that they do not obey Joy's law and have a much smaller footpoint separation. We confirm \citet{mun15} results that the sunspots appear to be two distinct populations.
\end{abstract}


\section{Introduction}

Magnetic fields generated at the base of the Sun's convective zone are thought to form toroidal flux tubes that become buoyant and rise to the surface \citep{par55, cha97, cha05}.  Sunspots often appear where the flux loops break the surface.  On average, bipolar sunspots show leading spots to be closer to the Equator than following spots.  \citet{hal19} first published observations of this phenomenon, now known as Joy's law, after statistical analysis showed that the mean tilt angle of bipolar sunspots increased with latitude. Joy's law has traditionally been interpreted as the Coriolis force operating on divergent plasma at the apex of a rising magnetic flux tube.  Rough calculations of the Coriolis effect on a rising flux tube by \citet{dsi93} measured a deflection in the tilt angle from an E-W orientation over time in terms of the rotational frequency of the Sun and emergence latitude.  The subsurface pitch angle of the toroidal field has also been proposed as a cause of tilt angle prior to the rise of flux tubes through the convection zone \citep{bab61}.

Numerical simulations of toroidal flux loops \citep{fan94} show a dependence of the tilt angle on the emerging latitude and field strength (B) where B $>$ 20 kG. Negative tilt angles start to occur when B $<$ 20 kG, where the weak field strength and low flux ($10^{20}$ Mx) host a converging flow at the apex in contrast to the standard divergent flow model \citep{web13}.  The thin flux tube approximation has been used to study rising magnetic loops in the convection zone and to explore the tilt angles and latitudes of emergence \citep{spr81, mor86}.  Studies show that the strength of the toroidal magnetic field should be around 30 - 100 kG in addition to indicating that the Coriolis force could explain the tilt angle described by Joy's law \citep{dsi93, cal95}. See \citet{fan09} for a review of these topics, including models of toroidal flux tubes rising with and without the influence of convective zone turbulence. \citet{web13} also compared flux tube simulations and found that tilt has a dependence on magnetic field strength in the flux tubes but no dependence on the total flux in the tube. 

Thin flux tube models that include convection and radiative diffusion predict shorter rise times through the convective zone \citep{web15}, producing tilt angles consistent with observed active regions. These simulations are able to reproduce tilt angles consistent with Joy's law without the need for anchored footpoints in the overshoot region.  However, tilt angle scatter for magnetic field strengths $\le 40$ kG are higher than observations.  

\citet{fis95} found that tilt was proportional to latitude and magnetic flux in the tube and inversely related to the magnetic field strength in the initial toroidal flux tube at the base of the convection zone. This version of Joy's law includes flux and initial field strength because aerodynamic drag balances the Coriolis force for rapidly emerging regions and magnetic tension balances the Coriolis force for slowly emerging regions. However, \citet{kos08} studied MDI magnetograms and found no evidence of a Joy's law dependence on magnetic flux or for a relaxation of the tilt angle toward zero after emergence.  Introducing magnetic field line twist stabilizes the cohesion of the rising tubes but also affects the tilt angle \citep{fan08}.  Since sunspot area and sunspot flux are highly correlated, area is used as a proxy for flux in order to study the relationship between the tilt angle and flux. \citet{jia14} binned Kodaikanal and Mt. Wilson Observatory tilt angle data according to sunspot size and found a weak correlation between mean tilt angle and sunspot group size, while the standard deviations significantly decrease with sunspot group size.

The amount of scatter in the tilt angles of bipolar regions provides information about the magnetic structures that produce sunspots.   \citet{wan89} analyzed 2710 bipolar magnetic regions (BMRs) and found no noticeable dependence of tilt angle on flux but higher deviations from the mean tilt angle for weaker BMRs. \citet{lon02} explained the departures of tilt angle from Joy's law during emergence as caused by upper convective zone turbulence.  Tilt angle scatter about Joy's law introduced by convection increases as flux decreases in thin flux tube models by \citet{web13}. \citet{ill15} showed the significant scatter in bipolar regions with areas less than 300 MSH (including ephemeral regions without sunspot activity).  A transition occurs between 300 and 400 MSH where the distribution of larger regions becomes dominated by sunspot activity with substantially less tilt angle scatter and more tilt angles that follow Joy's law.

 It has been proposed that the magnetic field could become so weak below the surface that convective zone turbulence dominates the flux tube.  \citet{fan94} suggested a mechanism of dynamic disconnection where a submerged portion of an emerged flux tube collapses after achieving hydrostatic equilibrium, disconnecting the tilt angle from the influence of the initial toroidal field pitch angle.  \citet{sch99} proposed subsurface reconnection of the untethered legs of the flux tube at depth to explain surface activity during decay.  \citet{lon02} rule out dynamic disconnection at shallow depths in their model of rising flux tubes.  \citet{sch05} modify the dynamic disconnection model based on strong, buoyancy-driven upflow and radiative cooling in a rising flux loop prior to emergence, finding disconnection depths around 5 Mm and disconnection times less than 3 days.  

As a flux loop ends its emergence and the footpoints stop separating, as observed in the photosphere as the centroid locations of the two polarities, Coriolis forces should end and the tilt angles should relax toward zero due to magnetic tension restoring field lines to an E-W orientation.  If the field is significantly frozen into the plasma and the differential rotation force is not strong enough, then the region maintains the tilt angle established prior to the end of emergence. The higher cadence of recently collected sunspot data presents a more complete picture of the tilt angle and footpoint separation over the lifetime of active regions.  A limited study of six active regions by \citet{pev03} using Solar and Heliospheric Observatory (SOHO)-MDI data reported separation distances around 25 Mm but generally increasing over time.  \citet{fan09} discussed the post-emergence evolution of subsurface fields (section 8.3) and observed that the photospheric portion of the footpoints stop separating at 100 Mm which cannot be explained in the $\Omega$-loop model.

In order to better understand the complexities of Joy's law, it is helpful to keep the following in mind.

\begin{enumerate}
\item  The pitch angle of the toroidal field beneath the surface may be a cause of the tilt angle prior to the rise of flux tubes through the convection zone.

\item Coriolis forces act on flows from the expanding plasma in the apex of the flux tubes rising through the bulk of the convection zone.  Coriolis forces increase with latitude and conversely should decrease near the Equator.

\item The high scatter in the tilt angle is attributed to the interaction of a rising flux tube with convection.  After emergence, subsurface convection should no longer impart scatter in tilt over the lifetime of the region. 

\item A disconnection of the flux tube from the source field would cause the tilt angle to no longer be affected by the intial pitch angle but instead relax to the angle held by the legs at the disconnection depth.

\end{enumerate}

\newpage
\section{Data}

Images taken by the Helioseismic and Magnetic Imager on board NASA's Solar Dynamics Observatory were used to calculate bipolar sunspot tilt angles for HMI - Debrecen Data (HMIDD)\footnote{$http://fenyi.solarobs.unideb.hu/ESA/HMIDD.html$} from 2010 April 30 to the present.  The calculation methods employed were an extension of those used by \citet{gyo11} on the SOHO/MDI - Debrecen Data (SDD). After correcting for limb darkening and flat field effects, Sunspot Automatic Measurement (SAM) software determined the penumbra borders from the first contour having a local maximum in an averaged gradient along contour (AGAC) with the umbra border contour having the global maximum in the AGAC.  Umbral area is defined by the number of pixels within the umbral border and reported in the HMIDD as millionths of solar hemispheres (MSH).  Within the umbra border, the centroid of the pixels weighted by intensity determined the umbral latitude and longitude of the spots.

Line-of-sight magnetic field information and umbral area measurements were used to calculate the mean latitude and longitude of the leading and following sunspot groups.  The HMIDD data used the polarity of sunspots only to separate the following and leading groups and do not indicate whether Hale's polarity rule is observed or not. After grouping sunspots by polarity, only longitude determines which group is considered the leading group.\footnote{$ftp://fenyi.solarobs.unideb.hu/pub/SDO/additional/tilt\_angle/Readme.txt$}  The HMIDD calculation of the tilt angle included the area-weighted latitude, area-weighted longitude, and latitude of the centroid of the entire bipolar region \citep[Eq. 1]{bar15}.  The latitude and longitude of the leading and following spot groups were determined by averaging the positions of all of the individual spots (weighted by area) within their respective group. We calculate the separation in degrees from the latitude and longitude of the leading and following spot groups and convert to Mm by equating 1\degr\ in separation with 12.13 Mm on the solar surface.

HMIDD data report tilt angles as positive in either hemisphere if the leading spot group is closer to the Equator than the following group.  Joy's law would be observed in the tilt angle as a function of the unsigned latitude.  However, we emphasize that near-Equator measurements of the tilt are incomplete since bipolar regions are assigned a hemisphere by latitude without regard to polarity \citep{mcc14}. At the time of publication, the HMIDD data only contained the beginning of Solar Cycle 24 when sunspot activity occurs at higher latitudes. 

\section{Bipolar Sunspot Behavior During Emergence and Decay}

Trends in tilt angle are difficult to observe in individual active regions and become more apparent in larger samples.  We identify 1151 NOAA regions in the HMIDD data and determine that 1111 regions had at least one umbral tilt angle reported. We use the umbral calculations of the latitude and longitude for the leading and following sunspot groups to determine footpoint separation, which we report as the distance between the centroids of the opposite polarities in the photosphere.  NOAA active regions often contain new emergence activity after previous umbral activity has stopped.  We exclude new activity in a particular NOAA region if the umbral activity was not reported for more than 24 hr. To minimize the foreshortening distortion of active regions observed near the limb, we limit the data to observations taken within 0.7 solar radii from the center of the Sun's disk. The hourly cadence of HMIDD data allow for binning of the tilt angle, total umbral area, and footpoint separation over 8 hr intervals for each active region. 

Individual active region information is recorded at various stages of development and decay depending on where the activity occurs on the solar surface in relation to the observation sight lines.  It would not be as useful to observe tilt angle behavior over the lifetime of an active region unless we calibrate the data to an active region characteristic that is observable in each region.  The onset of emergence would be an ideal reference point, but requiring regions to emerge on disk significantly limits the number of viable regions for study and emphasizes the emergence period over decay. The observational data of an active region that emerges on disk are more likely to exclude decay information, especially for longer-lived regions, as the regions rotate out of sight.   

The umbral area bin with the maximum value establishes the end of emergence and the beginning of sunspot decay for that region, creating a suitable common reference point across all of the active regions and placing equal emphasis on emergence and decay observations.   Data which start or end with peak umbral area for a region are excluded as these regions most likely began their decay prior to appearing on disk or did not complete their emergence period before vanishing off disk due to solar rotation away from the observational line of sight.  According to this restriction and all previously stated parameters, the number of viable regions to date available for study is limited to 235.

\citet{mun15} reconcile the area and flux distributions of the photospheric magnetic structures from multiple sunspot and active region databases into a composite of Weibull and log-normal distributions for flux below $10^{21}$ Mx and above $10^{22}$ Mx, respectively. They suggest that two separate mechanisms are ``giving rise to visible structures on the photosphere: one directly connected to the global component of the dynamo (and the generation of bipolar active regions), and the other with the small-scale component of the dynamo (and the fragmentation of magnetic structures due to their interaction with turbulent convection)'' \citet[p.18]{mun15}. Mu\~noz-Jaramillo et al. show that a shift in the HMI sunspot data from a Weibull distribution ($<10^{21}$ Mx) to a log-normal distribution ($>10^{22}$ Mx) occurs around 90 MSH in the umbral area. We use 90 MSH to separate tilt angle and footpoint separation data by peak umbral area into two data sets. A substantial number of regions fall below this threshhold, so we use 45 MSH to distinguish between small and midsize regions less than 90 MSH in peak umbral area.

\begin{figure}  [!ht]
\centerline{\includegraphics[width=0.99\textwidth,clip=]{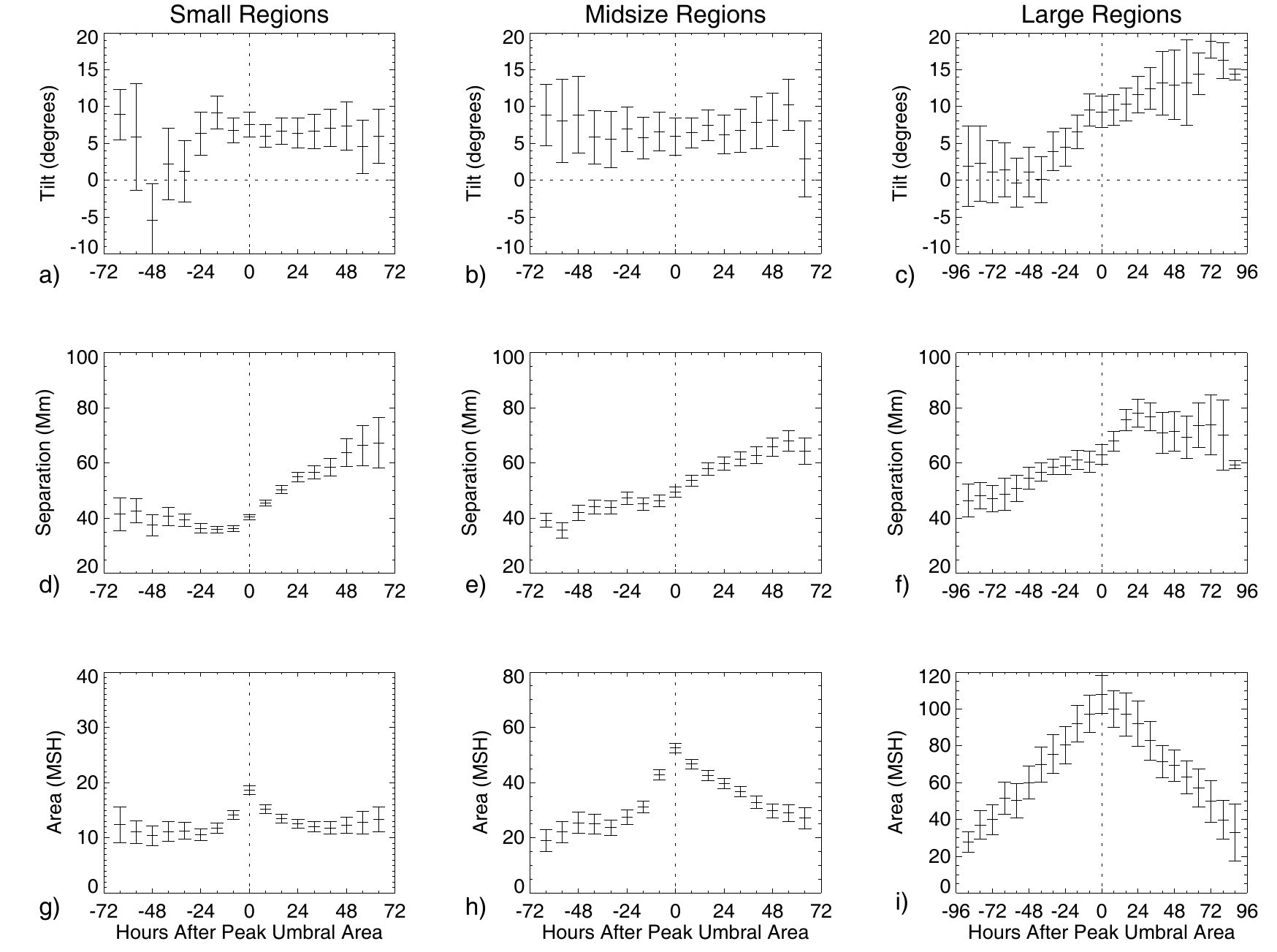}} 
\caption{Umbral tilt angle, footpoint separation, and total umbral area binned in 8 hr intervals are plotted in reference to the maximum umbral area ($UA_{max}$). Regions are separated into small ($UA_{max} <$ 45 MSH), midsize (45 $\le UA_{max} <$ 90 MSH), and large regions ($UA_{max} \ge$ 90 MSH). (a) Median tilt angle of small, (b) midsize, (c) and large regions. (d) Mean footpoint separation of small, (e) midsize, (f) and large regions. (g) Mean total umbral area for small, (h) midsize, (i) and large regions. Standard error of the mean overplotted as error bars.} \label{hmi9x}
\end{figure}

For each active region with a maximum umbral area ($UA_{max}$) of less than 45 MSH, we the sort tilt angle into 8 hr bins and find the median of each bin.  We do the same for the mean footpoint separation and mean umbral area. The median serves as a better measure of the center for low-sampled degree measurements that might include positive and negative values, however, the mean is preferred for non-negative measures of the separation and area. The time at which the peak umbral area is observed is noted such that all data before (after) that time are considered as emergence (decay).  The tilt, separation, and area values corresponding to the time of maximum umbral area are plotted at the t=0 point along the x-axis, see Figure \ref{hmi9x}.  We repeat the process to create two more data sets of midsize (45 $\le$ $UA_{max}$ $<$ 90 MSH) and large ($UA_{max}$ $\ge$ 90 MSH) regions.  Error bars are overplotted as the standard error of the mean. We excluded bins more than 72 hr (96 hr) from time zero for small and midsize (large) regions due to low sampling sizes at these times.

%

At the peak in umbral area, tilt angles are least noisy and typical for early cycle (higher latitude) activity.  Midsize regions have the lowest median tilt angle (5.9\degr) compared to small (7.6\degr) and large regions (9.3\degr), see Table \ref{areabin}.  Given the concentration of all of the regions at higher latitudes (Table \ref{areabin}), it is not useful to normalize the tilt angle for latitude until an entire solar cycle is observed. The dip in small region tilt angles around -48 hr (also visible in the separation and area plots) is likely an indication of regions that are smaller than average for this group beginning their emergence with negative tilt angles.  For two days before and after the peak area, the median tilt angles of midsize regions remain fairly constant, whereas the large regions show tilt angles occurring near zero earlier in the emergence period before steadily increasing over 4-5 days well into decay.     Higher variations are observed for all of the area classifications closer to the onset of emergence and toward the end of the decay period.  It is not unexpected to see more variability in the median tilt angles of small regions, especially at the beginning of their lifetimes when the influence of external factors such as convective turbulence is greater. Larger organized regions are more resistant to convective turbulence as seen in the more stabilized tilt angles.

Tilt angle increases in larger regions during emergence and decay but remains relatively steady for the small and midsize regions. We cite this as evidence of supergranular convective flows influencing the formation and evolution of smaller regions.  It may be that the evolution of larger regions is better explained by the $\Omega$-loop model, whereas smaller regions begin to form bipoles primarily from supergranular convection, although this does not preclude smaller regions from evolving beyond the supergranule model into larger regions.  This remains a topic of interest beyond the scope of our study.

%
\begin{table} [!ht]
\begin{center}
 \caption{Tilt Angle, Area, and Latitude at Maximum Umbral Area ($UA_{max}$)}\label{areabin}
 \begin{tabular}{l l l l}     
 \tableline\tableline
 & Small & Midsize & Large\\  
 & ($UA_{max} < 45$ MSH) & (45 $\le UA_{max} <$ 90 MSH) & ($UA_{max} \ge$ 90 MSH)\\
 \tableline
Median Tilt & 7.6\degr & 5.9\degr & 9.3\degr \\ 
Mean Area & 18 MSH & 52 MSH & 107 MSH \\ 
Median Latitude & 16.7\degr & 16.1\degr & 15.2\degr \\ 
n (viable) & 149 & 60 & 26\\ 
N (original) & 618 & 261 & 232\\  
 \tableline
 \end{tabular}
 \end{center}
\end{table}

All of the regions demonstrate a 3-4 day period of increase in footpoint separation with the onset of this period varying by region size.  Approximately 3 days before the umbral area peaks and lasting a day into decay, large regions increase in separation from around 45 to 75 Mm.  Midsize regions show a 30 Mm increase over 4 days as well, although the separation values are shifted downward slightly (40-70 Mm) and forward in time by about a day.  Small regions maintain 35-40 Mm in separation throughout the observed portion of the emergence period, then begin a 3 day period of increase in separation at the onset of decay, peaking near 70 Mm. At -16 hr, small regions organize around 35 Mm in separation with almost no variation.  This coincides with the beginning and end of two trends in separation for these regions: the onset of a steady increase in separation after a period of relatively constant mean separation values. 

The footpoint separation for small regions during the observed portion of the emergence phase remains relatively constant before beginning a period of separation at the start of decay, whereas the larger regions begin their separation prior to decay.  We suggest that these smaller regions have not yet accumulated enough flux to overcome the influences of supergranular convection before decay begins.  The divergent flow at the top of a supergranule cell pushes the flux to the cell boundaries as described by \citet{sch68} while the magnetic structure of the bipole begins to dictate the size and shape of the supergranule cell.  Smaller regions cannot begin their separation phase until the supergranular cell that aided in its formation begins to dissipate, typically after 1-2 days \citep{hir08}. We suggest that larger regions are less influenced by supergranular convection during the emergence phase and simply separate beyond the typical size of a supergranule cell as a result.

\begin{figure}[!htb]
    \centering
    \begin{minipage}{0.43\textwidth}
        \centering
        \includegraphics[width=0.99\linewidth, height=0.3\textheight]{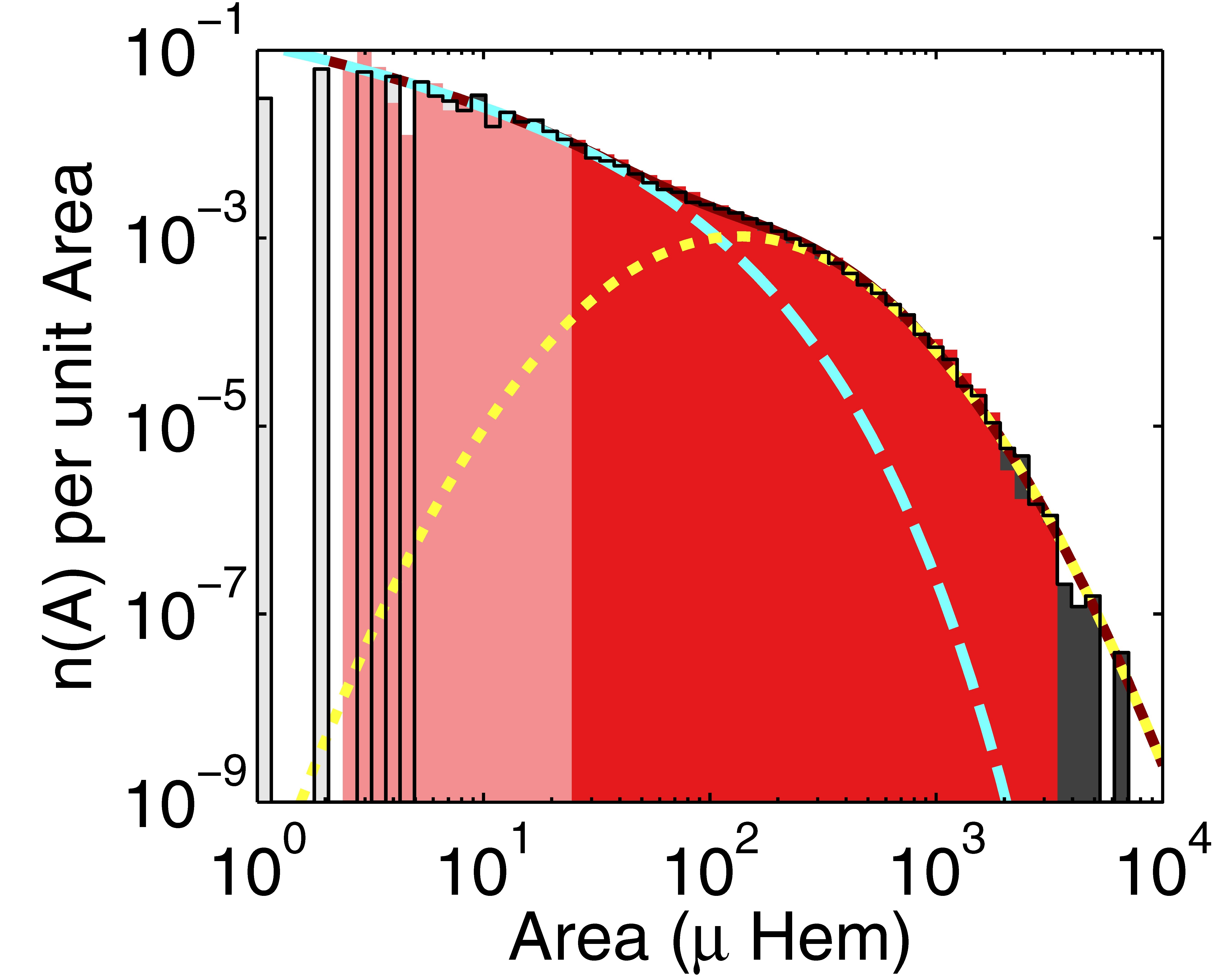}
        \caption{Empirical distribution of HMI sunspot areas (red), with Weibull (dashed blue line) and log-normal distributions (dotted yellow line) fitted to the darker red shade. Note the transition in distributions near 90 MSH. Reproduced with permission from \citet{mun15}.}
        \label{MUN15fig10g}
    \end{minipage}
\hfill
    \begin{minipage}{.53\textwidth}
        \centering
        \includegraphics[width=1.\linewidth, height=0.33\textheight]{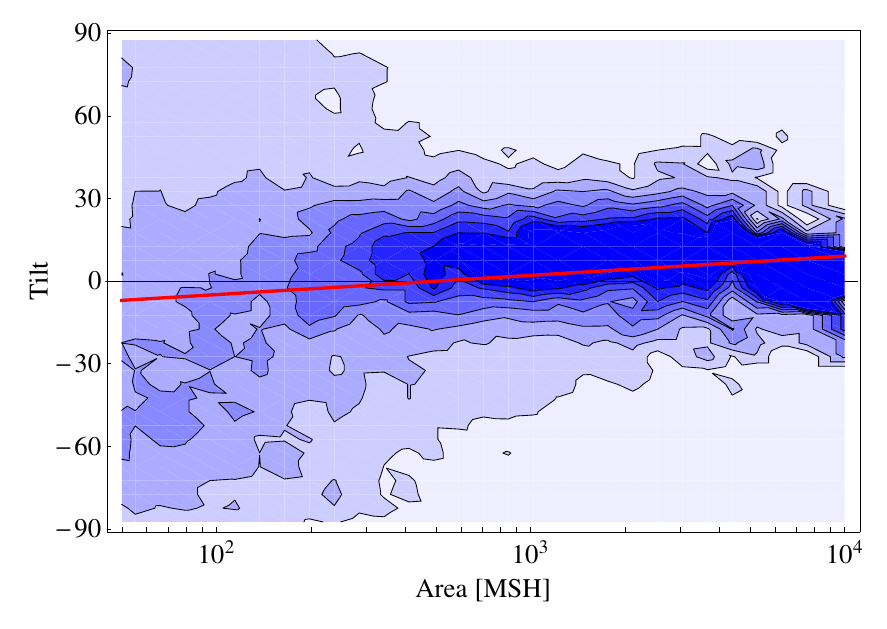}
        \caption{Distribution of BMR area and tilt angle from MDI data (1998-2007, latitudinal zone $| \theta| \ge 10 \degr $).  Color intensity indicates the number of bipoles relative to the total number of bipoles with the same areas.  The red line corresponds to the linear least-square fit.  Reproduced with permission from \citet{ill15}.}
        \label{ILLfig1}
    \end{minipage}
\end{figure}

The mean peak umbral areas for small regions ($\approx18$ MSH) and midsize regions ($\approx 52$ MSH) skew significantly lower in their respective area bins (Table \ref{areabin}).  \citet{mun15} determined that smaller sunspot regions ($<90$ MSH) display a Weibull distribution that also skews toward lower umbral areas. An empirical distribution of HMI data (Figure \ref{MUN15fig10g}) shows the transition at 90 MSH from the Weibull distribution of smaller regions to a log-normal distribution for larger regions. 

Including BMRs in the discussion extends the observation of tilt angle behavior to smaller regions that may or may not include sunspots.  It should be noted that BMR areas are reported to be up to 44 times larger than sunspot areas in the same active region \citep{cha11}.  Figure \ref{ILLfig1} shows a BMR distribution of the area and tilt angle from MDI data at higher latitudes ($| \theta| \ge 10 \degr $) where the color intensity indicates the number of bipoles relative to the total number of bipoles with the same areas. Two distinct distributions are visible due to the substantial amount of tilt angle scatter for BMRs less than 300 MSH as well as the trend toward negative tilt angles for smaller regions. 

\FloatBarrier

\section{Very Small Bipolar Sunspot Regions}

We have already seen evidence of smaller regions emerging with negative tilt angles (Figure \ref{hmi9x}a), particularly regions with shorter emergence periods and areas which are lower than average.  We limit our observations to very small peak umbral areas ($< 9$ MSH) in the HMIDD data.  Since these regions do not last as long, we are more likely to observe the entire life of the region within 0.7 solar radii from the center of the Sun's disk.  The behavior of very small active regions gives insight into how all active regions behave when they are first forming with smaller areas.  Due to the shorter lifetimes of these regions, we can apply the same binning techniques previously described but at a higher cadence of 2 hr.

\begin{figure}  [!ht]
\centerline{\includegraphics[width=0.4\textwidth,clip=]{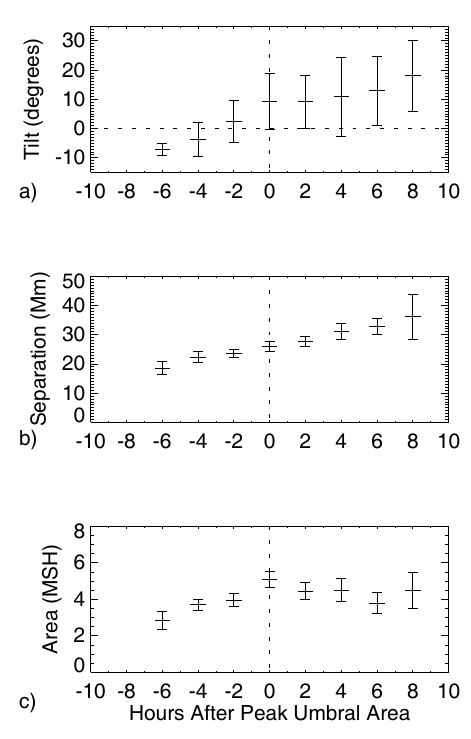}} 
\caption{(a) Median umbral tilt angle, (b) mean footpoint separation, and (c) total umbral area binned in 2 hr intervals are plotted in reference to the peak umbral area for regions of less than 9 MSH peak umbral area. Standard error of the mean overplotted as error bars.} \label{hmi3x}
\end{figure}

The median tilt angle, mean separation, and mean area are binned every 2 hr and calibrated to the time of peak umbral area according to previously described methods.  The standard error of the mean is overplotted as error bars for the tilt, separation, and area in Figure \ref{hmi3x}.  Given the previous parameters established for viable active regions as well as the size restriction to less than 9 MSH, our sample is small (n = 12) but still worthy of inclusion in the discussion.  Median tilt angles are anti-Joy (negative) at first, increasing to expected values in the decay period for activity at a median latitude of 16.1\degr.  Footpoint separation increases from 20 Mm to about 35 Mm in a 14 hr span ($\approx 298$ m s$^{-1}$), which is more rapidly than larger regions during the previously observed 3-4 periods of steady increase ($\approx 87$ m s$^{-1}$). The umbral area averages between 3 and 5 MSH throughout the lifetimes of these regions. 


Using MDI magnetograms, \citet{tla13} noted the anti-Joy behavior (negative tilt angles) of BMRs with areas less than 300 MSH at higher latitudes that was not observed in larger regions. \citet{cha11} report an average facular-to-sunspot ratio of approximately 44 to 1.  As a rough conversion of total umbral area to BMR area, 5 MSH in the umbral area equates to 220 MSH in the BMR area, which is less than the 300 MSH threshhold set by \citet{tla13}.  This permits a comparison of tilt angle results for our respective observations of smaller regions.  It appears that anti-Joy tilt angles occur in smaller regions during the emergence period and then rotate toward positive values as the decay period begins. Coriolis forces from the divergent plasma flow in the apex of a rising flux tube deflect E-W oriented bipoles toward the positive mean tilt angles typically reported. In contrast, any existing convergent flows would deflect toward negative tilt angles. The apex of a flux tube nearing emergence in the upper convective zone with weak toroidal field strength (15-30 kG) and low flux ($10^{20}$ Mx) hosts a converging flow \citep{fan94, web13}, producing tilt away from expected Joy's law angles.

\FloatBarrier

\section{Discussion}

When the peak umbral area is used to sort sunspot regions by size and to define the emergence and decay periods for each region, evidence of two size distributions emerges in the tilt angle behavior. After a period of near-zero tilt angles during emergence, a consistent increase in the large region ($UA_{max}$ $\ge 90$ MSH) tilt angle begins 48 hr before peak umbral area and lasts several days into the decay period. Regions smaller than 90 MSH show more consistency in tilt during this same time frame. It may be that larger regions eventually accumulate enough flux to form a new dynamic, with Coriolis forces acting on plasma draining from the apex of a more organized flux tube.  At some point during the accumulation of flux as larger regions emerge, Coriolis force induced tilt overtakes the influence of toroidal fields oriented in the E-W direction such that the increase in tilt angle persists into the decay period. 

We attribute the negative tilt angles in small regions ($UA_{max} < 45$ MSH) at -48 hr as evidence of new activity emerging with weaker initial field strengths \citep{web13} and consistent with other observational studies of smaller active regions \citep{tla13,ill15}. Our preliminary observations of very small regions ($UA_{max} < 9$ MSH) at a higher cadence indicate the presence of negative tilt angles during emergence. As the current solar cycle progresses, further study of small regions in the HMIDD data will be of interest.

A sustained period of increase in footpoint separation lasting several days occurs for all regions, but the onset of that period varies with region size. Small regions ($UA_{max}< 45$ MSH) do not increase in separation until the end of the emergence period, whereas larger regions begin several days earlier. Separation distances during the observed portion of the emergence period for small regions are consistently near 35-40 Mm, with almost no variation in the binned values near the peak in umbral area.  We attribute this behavior to the influence of supergranular convective cells as suggested by \citet{sch68} and discussed below. These small regions then steadily increase in separation during the first three days of decay, separating to at least 70 Mm. Midsize ($45 \le UA_{max}< 90$ MSH) and large regions ($UA_{max}\ge 90$ MSH) were observed to steadily increase in separation from 40 to 70 Mm and 45 to 75 Mm, respectively, from the start of the observed emergence period and lasting into decay. It should be noted that our observations of bipolar sunspot separation distances early in the emergence period are approximately 40 Mm, which are larger than the $\approx25$ Mm distances reported by \citet{pev03} for BMRs. This can be attributed in part to magnetic bipoles forming before the umbral intensity signature necessary for measuring bipolar sunspot separation distances is reached.

Dynamo theories of the solar magnetic cycle use the pitch angle of subsurface toroidal fields as a source of the initial tilt in active regions prior to emergence \citep{bab61, lei69}, however, pitch angle is not sufficient to explain all tilt angle behavior on the surface. \citet{sch68} suggests that the Coriolis forces from a divergent supergranular flow influences the tilt angle. He states that ``the horizontal Coriolis force needs a day of unimpeded uniform motion to rotate the resulting displacement by 6 degrees. Such motion can be found in a supergranulation cell which seems to last for 1 day (cf. Simon and Leighton, 1964). The same timescale holds for the appearance of a new active center" \citet[pp. 96-97]{sch68}. In modeling magnetic activity in the convective zone, \citet{wei71} considered Coriolis-induced cyclonic motion formed by the divergent flow at the top of convection cells, also noting the concentration of magnetic flux around the perimeters of the convection cells. The supergranule divergent flow and the resultant anti-cyclonic motion push magnetic flux to the boundaries of the cell and build up tilt angles as long as the flow persists. The footpoint separation sizes of very small regions or regions early in the emergence period are reasonably comparable to the observed distribution of supergranule diameter sizes \citep{hir08}. The separation velocities of the bipolar portion of very small regions are consistent with observed supergranular divergent flows of around 300 m s$^{-1}$ \citep{rie10, lan15}. It is likely that many active regions begin their emergence period with small umbral areas and that supergranular divergent flows influence the initial tilt angle and separation distances in these just forming regions. 

We cite the distinct variations in tilt angle and footpoint separation behaviors by region size as confirmation that two distributions of sunspot sizes exist. In order for a rising flux tube model to adequately describe all of the magnetic activity as it appears on the surface, some artificial assumptions are necessary. As more detailed observational data of magnetic regions become available, questions arise about the flux tube model. \citet{get15} discuss difficulties in comparing rising-tube model to observations and how a convective mechanism using in situ amplification and structuring of magnetic fields by convection avoids these difficulties. The rising flux tube model may be sufficient to describe the formation of larger bipolar sunspot regions, whereas the initial tilt angle and footpoint separation of smaller regions are heavily influenced by supergranular convection.  Further research into the cause of tilt angles in either sunspot size distribution is worth pursuing, especially the probability that a portion of the larger region distribution may be the result of smaller regions accumulating enough flux to expand beyond the influence of supergranular convection.  

\FloatBarrier



\begin{thebibliography}

\bibitem[Babcock(1961)]{bab61}
Babcock, H.W. 1961, \apj, 133, 572

\bibitem[Baranyi(2015)]{bar15}
Baranyi, T. 2015, \mnras, 447, 4 

\bibitem[Caligari et al.(1995)]{cal95}
Caligari, P., Moreno-Insertis, F., \& Sch\"{u}ssler, M. 1995, \apj, 441, 886 

\bibitem[Chapman et al.(2011)]{cha11}
Chapman, G.A., Dobias, J.J., \& Arias, T. 2011, \apj, 782, 150 

\bibitem[Charbonneau(2005)]{cha05}
Charbonneau, P. 2005, Liv. Rev. Solar Phys., 2, 2

\bibitem[Charbonneau \& MacGregor(1997)]{cha97}
Charbonneau, P. \& MacGregor, K.B. 1997, \apj, 486, 502

\bibitem[D'Silva \& Choudhuri(1993)]{dsi93}
D'Silva, S., \& Choudhuri, A. 1993, \aap, 272, 621

\bibitem[Fan(2008)]{fan08}
Fan, Y. 2008, \apj, 676, 680

\bibitem[Fan(2009)]{fan09}
Fan, Y. 2009, Liv. Rev. Solar Phys., 6, 4

\bibitem[Fan et al.(1994)]{fan94}
Fan, Y., Fisher, G. H., \& McClymont, A.N. 1994, \apj, 436, 907

\bibitem[Fisher et al.(1995)]{fis95}
Fisher, G.H., Fan, Y., \& Howard, R.F. 1995, \apj, 438, 463

\bibitem[Getling et al.(2015)]{get15}
Getling, A.V., Ishikawa, R., \& Buchnev, A.A. 2015, Advances in Space Research, 55, 862

\bibitem[Gy\H{o}ri et al.(2011)]{gyo11}
Gy\H{o}ri, L., Baranyi, T., \& Ludm{\'a}ny, A. 2011, in IAU Symp. 273, The Physics of Sun and Star Spots, ed. D. P. Choudhary \& K. G. Strassmeier (Cambridge: Cambridge Univ. Press), 403  

\bibitem[Hale et al.(1919)]{hal19}
Hale, G.E., Ellerman, F., Nicholson, S.B., \& Joy, A.H. 1919, \apj, 49, 153

\bibitem[Hirzberger et al.(2008)]{hir08}
Hirzberger, J., Gizon, L., Solanki, S.K., \& Duvall, T.L. 2008, \solphys, 251, 417

\bibitem[Illarionov et al.(2015)]{ill15}
Illarionov, E.,Tlatov, A.,\& Sokoloff, D. 2015, \solphys, 290, 351

\bibitem[Jiang et al.(2014)]{jia14}
Jiang, J., Cameron, R. H., \& Sch\"{u}ssler, M. 2014, \apj, 791, 5

\bibitem[Kosovichev \& Stenflo(2008)]{kos08}
Kosovichev, A.G., \& Stenflo, J.O. 2008, \apjl, 688, L115

\bibitem[Langfellner et al.(2015)]{lan15}
Langfellner, J., Gizon, L., \& Birch, A.C. 2015, \aap, 581, A67

\bibitem[Leighton(1969)]{lei69}
Leighton, R.B. 1969, \apj, 156, 1

\bibitem[Longcope \& Choudhuri(2002)]{lon02}
Longcope, D., \& Choudhuri, A. 2002, \solphys, 205, 63

\bibitem[McClintock \& Norton(2014)]{mcc14}
McClintock, B.H., \& Norton, A.A., Li, J. 2014, \apj, 797, 130

\bibitem[Moreno-Insertis(1986)]{mor86}
Moreno-Insertis, F. 1986, \apj, 166, 291 

\bibitem[Mu\~noz-Jaramillo et al.(2015)]{mun15}
Mu\~noz-Jaramillo, A., Senkpeil, R.R., Windmueller, J.C., et al. 2015, \apj, 800, 48

\bibitem[Parker(1955)]{par55}
Parker, E.N. 1955, \apj, 121, 491

\bibitem[Pevtsov et al.(2003)]{pev03}
Pevtsov, A.A., Maleev, V.M., \& Longcope, D.W. 2003, \apj, 593, 1217

\bibitem[Rieutord \& Rincon(2010)]{rie10}
Rieutord, M., \& Rincon, F. 2010, Living Reviews in Solar Physics, 7, 2

\bibitem[Schmidt(1968)]{sch68}
Schmidt, H.U. 1968, in IAU Symp. 35, Structure and Development of Solar Active Regions, ed. K.O. Kiepenheuer (Dordrecht: Reidel), 95

\bibitem[Schrijver \& Title(1999)]{sch99}
Schrijver, C.J. \& Title, A.M. 1999, \solphys, 188, 331 

\bibitem[Sch\"{u}ssler \& Rempel(2005)]{sch05}
Sch\"{u}ssler, M. \& Rempel, M. 2005, \aap, 441, 337 

\bibitem[Spruit(1981)]{spr81}
Spruit, H.C. 1981, \aap, 98, 155 

\bibitem[Tlatov et al.(2013)]{tla13}
Tlatov, A., Illarionov, E., Sokoloff, D., \& Pipin, V. 2013, \mnras, 432, 2975

\bibitem[Wang \& Sheeley(1989)]{wan89}
Wang, Y.-M., \& Sheeley, N.R. 1989, \solphys, 124, 81

\bibitem[Wang \& Sheeley(1991)]{wan91}
Wang, Y.-M., \& Sheeley, N.R. 1991, \apj, 375, 761

\bibitem[Weber \& Fan(2015)]{web15}
Weber, M.A., \&  Fan, Y. 2015, \solphys, 290, 1295 

\bibitem[Weber et al.(2013)]{web13}
Weber, M.A., Fan, Y., \& Miesch, M.S. 2013, \solphys, 287, 239

\bibitem[Weiss(1971)]{wei71}
Weiss, N.O. 1971, in IAU Symp. 43, Solar Magnetic Fields, ed. R. Howard (Dordrecht: Reidel), 757

\end{thebibliography}
\end{document}